\documentclass[12pt]{article}
\def\be{\begin{equation}}
\def\ee{\end{equation}}
\def\bea{\begin{eqnarray}}
\def\eea{\end{eqnarray}}
\usepackage{graphicx}

\catcode`\@=11
\def\lsim{\mathrel{\mathpalette\@versim<}}
\def\gsim{\mathrel{\mathpalette\@versim>}}
\def\@versim#1#2{\vcenter{\offinterlineskip
\ialign{$\m@th#1\hfil##\hfil$\crcr#2\crcr\sim\crcr } }}
\catcode`\@=12

\catcode`@=12
\begin{document}

\thispagestyle{empty}
\begin{flushright}
UCRHEP-T510\\
September 2011\
\end{flushright}
\vspace{0.3in}
\begin{center}
{\LARGE \bf Hiding the Higgs Boson from Prying Eyes\\}
\vspace{1.5in}
{\bf Ernest Ma\\}
\vspace{0.2in}
{\sl Department of Physics and Astronomy, University of California,\\ 
Riverside, California 92521, USA\\}
\end{center}
\vspace{1.5in}
\begin{abstract}\
There are two simple ways that the Higgs boson $H$ of the Standard Model (SM) 
may be more difficult to observe than expected at the Large Hadron 
Collider (LHC) or the Tevatron.  One is well-known, i.e. $H$ decays 
invisibly, into dark-matter scalar particles for example.  The other 
is that $H$ mixes with a heavy singlet scalar $S$ which couples to new 
colored fermions and scalars.  Of the two mass eigenstates, the light one 
could (accidentally) have a supppressed effective coupling to two gluons, 
and the heavy one could be kinematically beyond the reach of the LHC.
\end{abstract}

\newpage
\baselineskip 24pt

The one Higgs boson~\cite{d08} $H$ of the Standard Model (SM) of particle 
interactions is expected to be produced by gluon-gluon fusion at the Large 
Hadron Collider (LHC) and be observed through its decay into $ZZ$, $WW$, 
and other channels.  Recent reported data~\cite{atlas11,cms11,teva11} have 
excluded the following mass ranges at 95\% confidence level:
\begin{eqnarray}
{\rm ATLAS}: && 146-232,~~256-282,~~296-466~{\rm GeV}, \\ 
{\rm CMS}: && 145-216,~~226-288,~~310-340~{\rm GeV}, \\ 
{\rm TEVATRON}: && 156-177~{\rm GeV}.
\end{eqnarray}
Combined with the LEPII bound~\cite{pdg10} of $m_H > 114.4$ GeV, this 
leaves only a small window for its observation.  Whereas more data could 
eventually find $H$ through its rare decay mode to two photons, 
it is perhaps a good time now to consider how $H$ may be hidden from 
view because of either its decay or its production.

A first possibility is that $H$ decays significantly into invisible 
channels, thereby diminishing its branching fractions into observable 
final states.  This is a very old idea~\cite{ss82} and has many different 
model realizations.  One recent example is the model of a dark (inert) 
scalar doublet~\cite{m06}, where the Standard Model is extended to include 
a second scalar doublet, which is odd under an exactly conserved $Z_2$ 
symmetry~\cite{dm78}.  If the neutral member $\eta^0 = (\eta_R + i \eta_I)/
\sqrt{2}$ of this doublet is split so that $m_R < m_I$ by 
at least the order of 100 keV, then $\eta_R$ is a good dark-matter 
candidate~\cite{bhr06,lnot07}.  For the latest discussion on this model, 
see Ref.~\cite{ly11}.  If $2 m_R < m_H$, then the invisible decay 
of $H$ into these dark-matter scalars will suppress its branching fractions 
to other particles, as already discussed in detail a few years 
ago~\cite{cmr07}.  The effect is especially significant below the $WW$ 
threshold and could suppress the $\gamma \gamma$ branching fraction by 
as much as a factor of three.

A second possibility is a new proposal.  The idea is very simple.  
Suppose there is a scalar singlet $S$ which couples to new colored fermions 
and scalars.  In that case, both $H$ and $S$ will couple to two gluons 
through loops.  Let the $Hgg$ amplitude be $A_H$ and the $Sgg$ amplitude 
be $A_S$, then $A_H$ is dominated by the $t$-quark loop, and $A_S$ comes 
from the new colored fermions and scalars.  Take for example $A_S = 
3 A_H$.  Now if $H$ mixes with $S$, the linear combination 
$H' = (3H - S)/\sqrt{10}$ would not couple to two gluons, and would not be 
produced at the LHC by gluon-gluon fusion.  If $H'$ also happens to be a 
mass eigenstate, then it could hide from being seen at the LHC even if its 
mass is 170 GeV (above the $WW$ threshold).  The orthogonal combination 
$S' = (3S+H)/\sqrt{10}$ has an enhanced coupling to two gluons, but it is 
presumably heavy because it is mostly a singlet, and could be kinematically 
beyond the reach of the LHC.

Consider the scalar potential of the SM doublet $\Phi = (\phi^+,\phi^0)$ 
and a real singlet $S$:
\begin{equation}
V = \mu_1^2 \Phi^\dagger \Phi + {1 \over 2} \lambda_1 (\Phi^\dagger \Phi)^2 
+ {1 \over 2} \mu_2^2 S^2 + {1 \over 3} \mu_3 S^3 + {1 \over 4} \lambda_2 S^4 
+ \mu_4 S \Phi^\dagger \Phi + {1 \over 2} \lambda_3 S^2 \Phi^\dagger \Phi.
\end{equation}
Let $\langle \phi^0 \rangle = v$ and $\langle S \rangle = u$, then the minimum 
of $V$ is determined by
\begin{eqnarray}
0 &=& v (2 \mu_1^2 + 2 \lambda_1 v^2 + \lambda_3 u^2 + 2 \mu_4 u), \\ 
0 &=& u (\mu_2^2 + \lambda_2 u^2 + \lambda_3 v^2 + \mu_3 u) + \mu_4 v^2.
\end{eqnarray}
The $2 \times 2$ mass-squared matrix spanning the physical scalars $H$ and $S$ 
is given by
\begin{equation}
{\cal M}^2 = \pmatrix{ 2 \lambda_1 v^2 & \sqrt{2} (\lambda_3 u + \mu_4) v \cr 
\sqrt{2} (\lambda_3 u + \mu_4) v & 2 \lambda_2 u^2 + \mu_3 u - \mu_4 v^2/u}.
\end{equation}
Let the mass eigenstates of the above be $H' = H \cos \theta - S \sin \theta$ 
and $S' = S \cos \theta + H \sin \theta$, with eigenvalues $m_1^2$ and $m_2^2$, 
then Eq.~(7) may be rewritten as
\begin{equation}
{\cal M}^2 = \pmatrix{m_1^2 \cos^2 \theta + m_2^2 \sin^2 \theta & (m_2^2-m_1^2) 
\sin \theta \cos \theta \cr (m_2^2-m_1^2) \sin \theta \cos \theta & m_1^2 
\sin^2 \theta + m_2^2 \cos^2 \theta}.
\end{equation} 
As an example, let $\sin \theta = 1/\sqrt{10}$, $\cos \theta = 3/\sqrt{10}$, 
$u = 2 \sqrt{2} v = 492.4$ GeV, where $v = 174.1$ GeV, we then obtain 
$m_1 = 170$ GeV and $m_2 = 500$ GeV for the choice $\lambda_1 = 0.84$, 
$\lambda_2 = 0.47$, $\lambda_3 = 0.55$, and $\mu_3 = \mu_4 = 0$. This 
demonstrates the numerical viability of this proposal.

It has been assumed that $S$ couples to new colored fermions and scalars. 
This is of course model-dependent, but a necessary condition is to have 
$A_S = A(S \to g g)$ a few times larger than $A_H = A(H \to g g)$. 
Now $A_H$ is dominated by the $t$ quark which is a fundamental triplet 
under $SU(3)_C$ and is proportional to $(\sqrt{2} v)^{-1}$.  Suppose $A_S$ 
comes from a colored fermion octet $Q$ with the coupling $S \bar{Q} Q$, 
then it is proportional to $u^{-1}$ but its color factor of 3 is 6 times 
that of the $t$ quark.  Hence for the above choice of $u = 2 \sqrt{2} v$, 
$A_S \simeq 3 A_H$ is realized.  The allowed mass term $\bar{Q} Q$ would change 
the details of the above, but may be forbidden by a $Z_2$ symmetry under 
which $S$ and $Q_L$ are odd, but $Q_R$ is even.

More realistically, $H'$ is unlikely to decouple from $g g$ entirely. 
In that case, the suppression (or enhancement if $\sin \theta < 0$) factor 
in $H'$ production at the LHC is 
$(\cos \theta - (A_S/A_H) \sin \theta)^2$. On the other hand, depending 
on the choice of new colored fermions and scalars, there is also a 
contribution from $A(S \to \gamma \gamma)$ to $H'$ decay.  This means that 
the branching fraction of $H'$ to $\gamma \gamma$ would also not be the 
same as in the SM. 
If a particle is discovered at the LHC in the $\gamma \gamma$ channel below 
145 GeV, but with a branching fraction different from what is expected 
from the SM, especially if it is {\it greater}, 
it may be due to this effect.  The presence of the octet $Q$ may also be 
relevant in gauge-coupling unfication~\cite{m05} without supersymmetry.

In conclusion, the existence of the Higgs boson may be hidden from view at 
present because of a variety of scenarios, some of which have been discussed 
recently~\cite{bgow11,gms11,hv11,ht11}.  In this paper, two simple ways are 
considered: the presence of light dark-matter scalars which affects the 
decay or an accidental cancellation between $A(H \to gg)$ and $A(S \to gg)$ 
in $H-S$ mixing which affects the production.  In the latter case, 
an increase from the present $E_{cm} = 7$ TeV to 14 TeV at the LHC in the 
future would produce $S'$ easily, and the decay $S' \to H' H'$ would be a 
spectacular signature for discovering $H'$.

I thank Bohdan Grzadkowski and Maria Krawczyk for a stimulating 
``Scalars 2011'' Conference in Warsaw (August 2011) which led to this 
work. My research is supported in part by the U.~S.~Department of Energy 
under Grant No.~DE-AC02-06CH11357.

\newpage
\baselineskip 16pt
\bibliographystyle{unsrt}

\end{document}